\documentclass[twopage,11pt] {article}  

\newcommand{\lambdabar}{{\mathchar'26\mkern-9mu\lambda}}

\newcommand{\beq}{\begin{equation}}
\newcommand{\eeq}{\end{equation}}
\newcommand{\beqa}{\begin{eqnarray}}
\newcommand{\eeqa}{\end{eqnarray}}

\def\fun#1#2{\lower3.6pt\vbox{\baselineskip0pt\lineskip.9pt
\ialign{$\mathsurround=0pt#1\hfil##\hfil$\crcr#2\crcr\sim\crcr}}}

\usepackage{epsfig}
\begin{document}           

\title{Prompt muon-induced fission: a sensitive probe for
       nuclear energy dissipation and fission dynamics}  
\author{Volker E. Oberacker and A. Sait Umar\\
Department of Physics \& Astronomy, Vanderbilt University, \\ 
Nashville, TN 37235, USA\\
Feodor F. Karpeshin\\
St. Petersburg State University, RU-198504 St. Petersburg, Russia \\
Dept. Physics, Univ. of Coimbra, P-3004-516 Coimbra, Portugal} 

\maketitle                 

\pagestyle{myheadings} 
\thispagestyle{plain}         
\markboth{Oberacker}{Muon-induced fission} 
\setcounter{page}{1}         


\begin{abstract}
Following the formation of an excited muonic atom, inner shell
transitions may proceed without photon emission by inverse internal
conversion, i.e. the muonic excitation energy is transferred to the
nucleus. In actinides, the $2p \rightarrow 1s$ and the $3d \rightarrow 1s$ muonic
transitions result in excitation of the nuclear giant dipole and giant
quadrupole resonances, respectively, which act as 
doorway states for fission. The nuclear excitation energy is typically
$6.5 - 10$ MeV. Because the muon lifetime is long compared
to the timescale of prompt nuclear fission, the motion of the muon in
the Coulomb field of the fissioning nucleus may be utilized to learn
about the dynamics of fission. If there is large friction between the
outer fission barrier and the scission point the muon will
remain in the lowest molecular energy level and emerge in the $1s$ bound
state of the heavy fission fragment. On the other hand, if friction is
small (i.e. the nuclear collective motion is fast) 
there is a nonvanishing probability that the muon may be promoted
to higher-lying molecular orbitals, e.g. the $2p\sigma$ level, from where
it will end up attached to the light fission fragment. Therefore,
theoretical studies of the muon-attachment probability to the light
fission fragment, $P_L$, in combination with experimental data can be
utilized to analyze the dynamics of fission, and nuclear energy
dissipation in particular. In this way, one may be able to distinguish
between nuclear energy dissipation arising from two-body collisions
and ``one-body friction'' caused by the mean field.
We study the dynamics of a muon bound to a fissioning
actinide nucleus in two different ways: first, we solve the non-relativistic
time-dependent Schr\"odinger equation using the Born-Oppenheimer expansion
method in terms of quasimolecular wavefunctions.
In the second approach, we solve the relativistic time-dependent Dirac
equation on a 3-D Cartesian lattice utilizing the B-Spline
collocation method. For $^{237}_{\ 93}$Np we find a dissipated energy of order
$0-10$ MeV and a fission time delay due to friction of up to $2 \cdot 10^{-21}$ s.
\end{abstract}


\section{Introduction}

Nuclear physics experiments with $\mu ^{-}$ beams provide information on
fundamental symmetries and interactions. Muonic atoms, in particular, have proven
extremely useful in examining the electromagnetic properties of nuclei, e.g.
electric charge distributions and multipole moments, because the muon has a
high position probability density inside the nucleus owing to its small
Compton wavelength $\lambda _c=\hbar /(m_\mu c)=1.87fm$ \cite{Ki71}.
After muons have been captured into high-lying single particle states
they form an excited muonic
atom. Inner shell transitions may proceed without photon emission by
inverse internal conversion \cite{knes}, i.e. the muonic excitation energy is
transferred to the nucleus. In actinides, the $E1: (2p \rightarrow 1s, 6.6 MeV)$ and the
$E2: (3d \rightarrow 1s, 9.5MeV)$ muonic transitions result in excitation of the nuclear
giant dipole and giant quadrupole resonances, respectively, which act as
doorway states for fission. Following these atomic transitions, 
the muon will be in the ground state of the actinide atom and can be utilized to probe
the fission dynamics. Finally though, the muon is going to be captured by
one of the fission fragments. However, nuclear capture as a result of the weak
interaction occurs on a time scale of order $10^{-7}$ s which is many orders
of magnitude larger than the time scale of fission.

From a theoretical point of view, prompt muon-induced fission has several
attractive features. Because the nuclear excitation energy equals or exceeds the
fission barrier height it is permissible to treat the fission
dynamics classically (no barrier tunneling). The muon dynamics is determined
by the electromagnetic interaction which is precisely known; hence, the
process can be calculated, at least in principle, with any desired
precision. Our main task is the solution of the Dirac equation for the muon
in the presence of a time-dependent external Coulomb field which is
generated by the fission fragments in motion. 

We will demonstrate that the
muon attachment to the light fission fragment depends on the nuclear
friction between the outer fission barrier and the scission point. In this
context, nuclear friction is defined as the irreversible flow of energy (and
linear or angular momentum) from collective to intrinsic single-particle
motion\cite{Ha78}. We include in our classical dynamical calculations for
the fission mode a linear friction force to account for energy dissipation
via neutron and photon emission. Through muon-induced fission one expects to
gain a deeper understanding of the energy dissipation mechanism in
large-amplitude nuclear collective motion. 

The prompt muon-induced fission process is most easily understood via a
``correlation diagram'', i.e. one plots the single-particle energies of the
transient muonic molecule as a function of the internuclear distance (see Fig.3 of
ref.\cite{OU93}). If there is a large amount of friction during the motion
from the outer fission barrier to the scission point the muon will
remain in the lowest molecular energy
level $1s\sigma$ and emerge in the $1s$ bound
state of the {\it heavy} fission fragment. If, on the other hand, friction is
small and hence the nuclear collective motion is relatively
fast there is a nonvanishing probability
that the muon may be promoted to higher-lying molecular orbitals, e.g.
the $2p\sigma$ level, from where it will end up attached to the {\it light}
fission fragment. Therefore, theoretical studies of the muon-attachment
probability to the light fission fragment, $P_L$, in combination with
experimental data can be utilized to analyze the dynamics of fission,
and nuclear energy dissipation in particular.

There are two different mechanisms that contribute to nuclear energy
dissipation: two-body collisions and ``one-body
friction''. The latter is caused by the moving walls of the
self-consistent nuclear mean field. The role played by these two
dissipation mechanisms in fission and heavy-ion reactions is not yet 
completely understood. In 1976, in a pioneering work  
Davies, Sierk and Nix \cite{DSN76} calculated the effect of
viscosity on the dynamics of fission. Assuming that friction is caused by
two-body collisions they extracted a viscosity coefficient $\mu = 0.015$
Tera Poise from a comparison of
theoretical and experimental values for the kinetic
energies of fission fragments. The corresponding time delay for the
nuclear motion from the saddle to the scission point was found to be of
order $\Delta t =1 \times 10^{-21}$ s. However, in one-body dissipation
models the time delay is an order of magnitude larger.

Theoretical studies in combination with rather scarce experimental
data on prompt fission obtained so far give rise to a
definite answer: Estimates of the energy  dissipated
in the case of the one-body mechanism were made in the two-center
harmonic oscillator model in analytical form \cite{BO2000}.
For this purpose, the Born-Oppenheimer expansion method
developed for calculation of the muon promotion probability,
was applied for assessment of the nucleon promotion rate to higher
quasimolecular orbitals, arising from the translation of the nascing
fragments. The last process provides the microscopic
mechanism for the one-body dissipation. The obtained estimate of the
dissipated energy, $E_{diss} < 1MeV$, allows 
one to rule out the one-body mechanism,
even with all the appoximations of the model kept in mind. 

Several experimental techniques are sensitive to the energy dissipation
in nuclear fission. At high excitation energy, the multiplicity of
pre-scission neutrons \cite{Ga87} or photons \cite{Ho95} depends on the
dissipation strength. At low excitation energy, the process of
prompt muon-induced fission \cite{MO80} provides a suitable ``clock''.
This process will be discussed here.

Nonrelativistic calculations for muon-induced fission have been carried out
by several theory groups, starting in 1980 \cite{MO80,MW80,ON80,MW81,Ka97}, but
relativistic calculations using the Dirac equation for the muon only became
feasible in 1992 \cite{OU92,OU93,OU98,O99}. We discuss here in detail our
theoretical approach and the numerical implementation, and we compare
our results with experimental data obtained at the Los Alamos Meson Physics
Facility (LAMPF) \cite{WJ78,SW79}, at the Tri-University Meson Facility
(TRIUMF) \cite{AB80,Ka80}, and at CERN and the Paul Scherrer Institute
(PSI) \cite{GH78,JK80,Po81,Po89,RB91} .


\section{Prompt and Delayed Fission Induced by Muons}

Following the irradiation of a target with a $\mu ^{-}$ beam the muons lose
most of their kinetic energy by ionization in the target material within $%
10^{-9}$ to $10^{-10}s$. Once their velocity has become comparable to
the orbital electron velocities characteristic of these atoms,
they are slowed down further by inelastic collisions with valence
electrons and are finally captured into high-lying states
($n_{\mu} \approx 14$)
forming a muonic atom. The theoretical aspects of the interaction of
muons with condensed matter were first studied by Fermi and Teller \cite
{FT47} and were later explored in more detail by Wu and Wilets \cite{WW69}
and by Kim \cite{Ki71}. Because all muonic bound states are unoccupied the
muon will cascade down to the ground state within $10^{-13}s$. From the
outer shells the excited muonic atom decays preferentially by emission of
Auger electrons. Since $\Delta E$ increases rapidly for the inner shells,
the transitions between levels with $n\le 5$ are dominated by mesic X-rays.
Alternatively, the transitions may proceed without emission of radiation via
inverse internal conversion. From the K-shell, the muon disappears at a
characteristic rate $\lambda =\lambda _0+\lambda _c$, where $\lambda
_0=(2.2\times 10^{-6}s)^{-1}$ denotes the free leptonic decay rate and $%
\lambda _c$ the nuclear capture rate; $\lambda _c$ depends upon the charge
and mass of the nucleus (Goulard-Primakoff formula \cite{GP74}) and is of
order $(7.5\times 10^{-8}s)^{-1}$ for actinides \cite{SW79}. Muons stopped
in an actinide target may induce nuclear fission in two different ways:

$\bullet $ {\it Delayed Fission Following Nuclear Muon Capture}

The muon is captured by a proton inside the nucleus and forms a neutron and
a muon neutrino

\begin{equation}
\mu ^{-}+(Z,A)\rightarrow (Z-1,A)^{*}+\nu _\mu .
\end{equation}

Even though most of the energy is taken away by the neutrino, the average
nuclear excitation energy is $15-20MeV$ which is well above the fission
barrier for actinides $E_f=5-6MeV$. Fission via nuclear muon capture is {\it %
delayed}, i.e. it occurs with the characteristic mean lifetime of the weak
decay process, $\tau _{capt}=(7-8)\times 10^{-8}s$.

$\bullet $ {\it Prompt Fission Resulting From Inverse Internal Conversion
In Muonic Atoms}

In this case, the excitation energy of the muonic atom is transferred to the
nucleus by an internal conversion process (nonradiative transition) and the
muon ends up in the K-shell of the muonic atom
\begin{equation}
(Z,A)(\mu ^{-})^{*}\rightarrow (Z,A)^{*}\mu ^{-}.
\end{equation}

For the innermost atomic transitions in an actinide muonic atom, the
transition energy generally exceeds the fission barrier height. The result
is {\it prompt fission in the presence of the muon}, since the muon is not
annihilated by this process, in contrast to fission resulting from nuclear
muon capture. The nucleus will be surrounded by the muon during the entire
fission process, unless the muon is ionized. Eventually, the muon will decay
by nuclear muon capture from the fission fragments. Experimentally, both
fission modes can be distinguished because of their different time scale. In
this paper, we focus on prompt muon-induced fission. This process was first
discussed by Wheeler \cite{Wh49} and considered in more detail by Zaretski
and Novikov \cite{ZN61}. It is important to know the specific atomic
transitions that are responsible for prompt fission. Fig.1 of ref.\cite{OU93} shows
the Coulomb interaction energy between the muon and a $^{238}$U nucleus as
well as the binding energies of the lowest bound states. Even though E0
transitions such as $2s\rightarrow 1s$ and $3s\rightarrow 1s$ exhibit the
largest internal conversion rates they do not contribute to fission because
they lead to excitation of the giant monopole resonance which is spherically
symmetric and much too high in energy. On the other hand, the $%
(E1:2p\rightarrow 1s)$ and the $(E2:3d\rightarrow 1s)$ transitions result in
excitation of the electric giant dipole and quadrupole resonances,
respectively, both of which act as doorway states for fission.
Let us consider the specific
case of $^{238}U$: The giant dipole resonance is located at $E_{GDR}=12.8MeV$
and has a width $\Gamma =6MeV$\cite{BF75}; for the $T=0$ giant quadrupole
resonance the corresponding numbers are $E_{GQR}=9.9MeV$ and $\Gamma =6.8MeV$%
\cite{Ne78}. According to Teller and Weiss \cite{TW79} and Karpeshin and
Nesterenko \cite{knes} it is very probable
that the $3d\rightarrow 1s$ radiationless transition will be dominant for
muon-induced fission, because its transition energy of $9.6MeV$ is very
close to the peak of the giant quadrupole resonance whereas the $%
2p\rightarrow 1s$ transition energy of $6.6MeV$ is far off the center of the
giant dipole resonance. Experimentally, the situation is controversial:
Johansson et al. \cite{JK80} measured muonic X-rays in coincidence with
prompt fission in $^{238}U.$ From the muonic X-ray intensity ratios for
prompt and delayed fission they conclude that $(74\pm 15)\%$ of all prompt
events can be attributed to the $3d\rightarrow 1s$ radiationless transition
and only $(26\pm 15)\%$ to the $2p\rightarrow 1s$ transition. On the other
hand, Kaplan et al. \cite{Ka80} find in similar studies a predominance of
the E1 transitions in their prompt fission data.

As mentioned in the Introduction, if the nuclear
motion is relatively fast (low friction) there is a nonvanishing probability
that the muon may be promoted to higher-lying molecular orbitals, e.g.
the $2p\sigma $ level from where it may end up attached to the light fission
fragment. This gave rise to hopes that the muon attachment probability to the light
fragment $P_L$ would depend on the pre-fission scenario. 
In this case, theoretical studies of the muon-attachment probability to
the light fission fragment in combination with experimental data can be
utilized to probe the dynamics of prompt fission. 

On the other hand, in a series of papers it was pointed out that the
fragment motion occurs quasiclassically. The fragment velocity exceeds
by $\sim$ 6 times that of the muon in the orbit. Moreover, it
satisfies the Massey criterium of adiabaticity which does not
promise any significant transition probability. 

Fortunately, the presence of the avoided crossing of the $1s\sigma$ and
$2p\sigma$ levels favors the transition. As is known, in this case the
transition occurs in a narrow vicinity of the crossing. At the same time,
the pseudocrossing occurs well beyond the scission point, at a distance
which is twice as large as that of the neck rupture for the most likely
pairs of fragments. At this distance, the trajectory of the relative
fragment motion $R(t)$ is governed by their mutual Coulomb repulsion.
Therefore, knowing the $P_L$ value might only provide us with the instantaneous
velocity of the fragments at this point, which could be obtained much more
easily from conservation of energy. By contrast, it could be argued
that the avoided crossing
approaches the point of rupture as the fragment asymmetry increases.
This suggests that the $P_L$ value also becomes more sensitive
to the pre-fission scenario in this case. 

Furthermore, in a series of papers \cite{jpg04,elseyaf}
it was noted that the point of rupture
itself can be considered as an irregular point where the
analyticity of the dependence of $R(t)$ on time is broken.
That is rather apparent from the mathematical viewpoint, as the
analyticity would assume a unique trajectory in all the domain of
its definition, $0\leq R < \infty$. Actually, the trajectory
can be considered unique only after scission, for $R_{sc} <R<\infty$.
But before scission, one can provide a rather arbitrary scenario,
as we saw when discussing the one- and two-body mechanisms of
friction. Moreover, the very position of the scission point itself
may be varied within certain limits. We return to this question
in a later section.

In fact, the muon appears
to be the only available tool for such studies. However, this simple picture
is complicated by the fact that transitions to some of the higher-lying
levels of the transient muonic molecule (e.g. $2p\pi $ and $2s\sigma $)
result again in muon attachment to the heavy fragment.

To obtain an order-of-magnitude estimate for the muon attachment
probabilities, we utilize a simple formula derived by Demkov \cite{De63} and
by Meyerhof \cite{Me73}. Their model is based on the two lowest molecular
levels (1s$\sigma $ and 2p$\sigma $) and utilizes first-order perturbation
theory to calculate the transition probability from the 1s$\sigma $ to the 2p%
$\sigma $ level; within the two-level model, it is equal to the muon
attachment probability to the light fission fragment:
\begin{equation}
P_L=\left( 1+e^{2\mid x\mid }\right) ^{-1},\bigskip\ x=\frac{\pi \left(
I_H-I_L\right) }{\left( \frac vc\right) \sqrt{2m_\mu c^2}\left( \sqrt{I_H}+
\sqrt{I_L}\right) },
\end{equation}
where I$_H$ and I$_L$ denote the binding energies of the muonic K-shell
belonging to the heavy and light fission fragments, respectively, and $v$ is
the relative velocity of the fission fragments. For a fragment charge
asymmetry $\xi =Z_H/Z_L=55/39=1.41$ (corresponding to the peak of the mass
distribution) one finds I$_H$=5.93 MeV and I$_L$=3.45 MeV. If we assume that
the molecular transition occurs at a relative velocity of the fission
fragments of $v=0.08c$ we obtain a muon attachment probability $P_L=0.042$
for the light fission fragment; this value decreases to $P_L=0.015$ if $%
v=0.06c.$ The choice of the parameters in the Demkov model and its extension
are discussed in more detail in the next section.


\section{Physical premises}

\subsection{Analytical properties of the trajectories}

The concept of determining the role of the avoided crossing for the transition probability is a milestone in the quasiclassical theory of collisions accompanied with reorganization of the scattering systems, in the case where one or two of the subsystems are described quasiclassically, and the other subsystems are described in terms of quantum mechanics. Most famous are collisions of the Landau-Zener type. Well-known are also the Demkov and Nikitin models.

These models are based on the different character of the energy term's behavior in the vicinity of the scission point. In the two-level approximation, the usual expression for the terms can be written as follows:
\beq
E_{1,2 } = \frac{1}{2}(\epsilon_1 + \epsilon_2) \pm 
\sqrt{(\epsilon_1 - \epsilon_2)^2 + \Delta^2} \;\;\;.
\label{LZ1}
\eeq

The physical meaning of the parameters $\epsilon_1$, $\epsilon_2$ in eq. 
(\ref{LZ1}) is that they are considered to be the non-interacting levels. The parameter $\Delta$ describes their interaction, 
and $E_1$, $E_2$ are the true terms, with the interaction taken into account. 

The Landau-Zener model deals with intersecting terms $\epsilon_1$, $\epsilon_2$. The 
interaction $\Delta$ is then considered to be $\Delta = constant $ in the vicinity of the crossing. 
In the opposite extreme, the Demkov model considers the case of $\epsilon_1$, $\epsilon_2$ parallel to 
each other, which is usually the case at large distances. At smaller distances, the terms start to diverge due to the interaction, which is supposed to be of exponential form: 
\beq
\epsilon_1- \epsilon_2 \equiv \epsilon = constant, \;\;\;\Delta = \exp (-\lambda R)\;\;.
\eeq

	The Nikitin model generalizes both the  Landau-Zener and Demkov models with the following parameters:
\beq
\epsilon_1- \epsilon_2 =  \epsilon - B \cos \theta \exp (-\lambda R)\;\;\;, 
\Delta = B \sin \theta \exp (-\lambda R)\;\;.
\eeq
Specifically, the Demkov model is obtained with $\theta = \pi / 2$.
The level crossing occurs in the complex plane at the point 
\beq
R_c = (\epsilon / B) \exp (i\theta)\;\;.
\eeq

Many characteristic features of the muon distribution were consecutively studied within the framework of the Demkov \cite{dem,kostr} and Nikitin \cite{kar82} models. 
The parameters of the models were related to realistic muonic wavefunctions. All the parameters, as well as the relative velocity, entering the expression for the transition probability, 
have  been taken at the point of the avoided crossing. Naturally, pre-fission dynamics was not involved in such an approach, as the pseudocrossing point is beyond the scission point as mentioned in the Introduction. 
Conclusions about the pre-fission dynamics could be drawn from comparison with experimental data (e.g., \cite{kar87}). Moreover, prospects of the study of other manifestations of the fission dynamics, such as muon shake-off due to the neck rupture, have been discussed. 
These effects are not in the scope of the quasimolecular adiabatic picture considered herein. The results obtained were also applied for the interpretation of many experiments concerning muonic conversion of the prompt fission fragment $\gamma$ rays, and muon capture by the fragments (e.g., \cite{Ka97} and refs. cited therein).

\subsection{The complex trajectory method}

What was stated previously about the analytical properties of the 
trajectory function $R(t)$ can be realized in terms of the 
complex trajectory method \cite{jpg04,elseyaf,elseizv}.

\protect
\begin{figure}[!hbt]
\centerline{ \epsfxsize=10cm \epsfysize=8cm \epsfbox{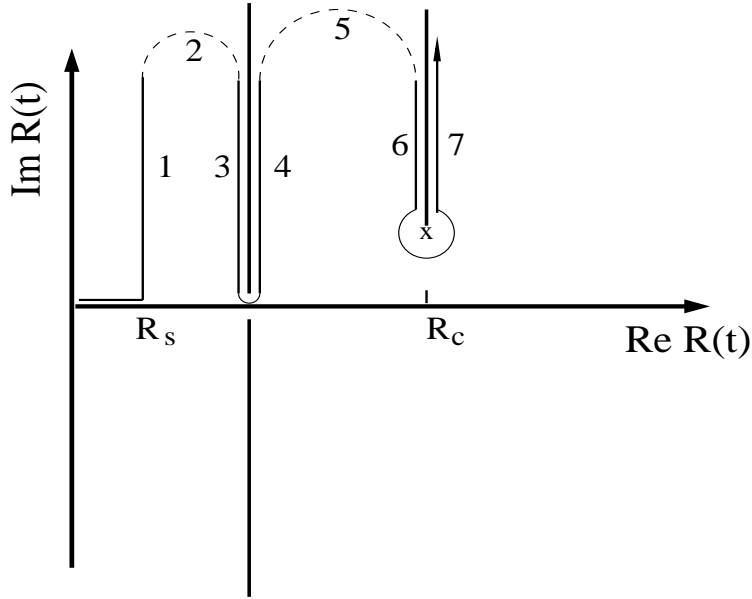} }
\caption{ \protect\footnotesize The complex trajectory. The point  $R = 0$
corresponds to the united atom limit. The onset of population of
the $2p\sigma$ level is at the saddle point $R_s$. The scission
point $R_{sc}$ and the point of avoided crossing $R_c$ are the other
singular points, where the integration contour is fixed.}\label{scf1}
\end{figure}

According to this method, the path $R(t)$ corresponding to the
trajectory of the relative motion of the fragments from $R = 0$ to
$R \to \infty$ can be deformed and shifted to the upper half-plane
of complex R values, with the  exception of irregular points.

The reason for such a transformation is that in the upper 
half-plçane the nonadiabatic matrix element vanishes. As a result, 
the main contribution comes from small segments nearest to the real 
axis of $R$, attached to the fixed irregular points. This 
transformation is shown in Figure \ref{scf1}. The irregular points 
are, as was mentioned previously, the pseudocrossing point $R_c$, 
which is a square-root type branching point, the scission point 
$R_{sc}$, and the starting point of the motion which in our case 
is reasonable to put on the top of the barrier. Thus,  the contributions from the
line segments 2 and 5 are expected to vanish for $R \to \infty$
\cite{land}.

    We then note that the contributions from the segments 3 and 
4 partly cancel one another,
being of the opposite sign. Were the scission point a regular
point of the trajectory, the cancellation would be complete, and
this point would give no contribution to the transition amplitude.
This would be the same as the point $R_{sc}$ would be shifted to
the upper half-plane, together with the segments 3 and 5. 
    The two vertical cuts in Figure \ref{scf1} fix the trajectory 
at the scission point,
not allowing one to move it from the real axis of $R$.

    Relative contributions of the scission and pseudocrossing points
turn out to be different, depending on the mass split of the
fragments. The first contribution is several times smaller than
the second one in the case of the most probable mass split. In the
case of larger asymmetry, the contribution from scission
approximately holds. In contrast, the contribution from the
avoided crossing exponentially decreases. For this reason, the
attachment probability becomes relatively more sensitive to the
pre-fission scenario for more asymmetric fission, though its
absolute value decreases. This circumstance was noted in refs.
\cite{ka92a,kar90}, where a higher sensitivity for larger
asymmetry was predicted as a consequence of approaching the
pseudocrossing point to the prefission area.

    We note that in actual calculations, the relative velocity 
$V(R)$ is smoothly joined at the point $R_{sc}$ from the left 
and the right, with its derivative. Then the position of $R_{sc}$ 
is the only parameter which governs the contribution from the 
scission point. In this schematic picture, the pseudocrossing point 
appears to be completely independent of the prefission scenario, as the 
values of integrands along the segments 6 and 7 are determined by 
analytical continuation of the Coulomb part of the trajectory at 
$R_{sc} < R < \infty$. This explains why in some papers the resulting transition 
probability was found not to depend on the prescission scenario: 
as a matter of fact, the position of $R_{sc}$ was not varied in 
those studies, but only the prefission time $t_{sc}$ \cite{fior,kar87}.

\subsection{The Born - Oppenheimer expansion}

Many of the first papers devoted to the calculation of
 the muon distribution were based on the Born - Oppenheimer 
expansion. We use the following series expansion \cite{BO2000}:
\begin{eqnarray}
\Psi (\mathbf {r}; R(t))= \sum_{n} C_n \Phi_n(\mathbf{ r}; R(t))
e^{i (\mathbf{p}_n \mathbf{r})} e^{-i \int^t E'_n (t') dt'} .
\label{sc1}
\end{eqnarray}
Here $\Psi (\mathbf {r}; R(t))$ is the wavefunction,
 $\Phi_n$ form a full set of quasimolecular wavefunctions.
Our basic product functions $\Phi_n(\mathbf{r}; R) e^{i
\mathbf{p_n r}}$ satisfy the time-dependent Schr\"{o}dinger
equation
\begin{eqnarray}
(i \frac{\partial}{\partial t} - H  ) \Phi_n(\mathbf{r}; R) e^{i
\mathbf{p}_n \mathbf{r} - i E_n' t} = 0  ,   \label{sc2}
\end{eqnarray}
with modified eigenvalue  $E'_n = E_n + p^2_n / 2 \mu$, reflecting
conservation of energy, and the two-centre Hamiltonian \beq H = -
{\frac{ \Delta }{2 \mu }} + V_1(|\mathbf{r - R}_1|)
 + V_2(|\mathbf{r - R}_2|)  .   \label{sc3}
\eeq

When the internuclear distance $R \to \infty$ each of the
functions $\Phi_n$  correlates with a certain muonic
state n of the corresponding fragment. \\
\hspace*{1em}   The momentum-translation exponents (m.t.e.) are
introduced in eq. (\ref{sc1}), where $\mathbf{p}_n = \mu \mathbf{
v}_n$, with $\mu$  the muon mass, and $v_n$ a constant, equal to
the velocity of the fragment in the  asymptotic region. This
condition is not essential, and can be easily abandoned
\cite{jpg04}. On the other hand, the orthogonality of the basis
functions is  broken in this description.

Substituting  the expansion (1) into the
time-dependent Schr\"{o}dinger equation, and neglecting  small
non-orthogonal terms, we get  the following set of coupled
equations
\begin{eqnarray}
\frac{dC_i}{dR} = - \sum_k F_{ik} C_k    ,   \label{sc4}   \\
F_{ik} = {\cal M}_{ik}
 e^{-i\int^{R} [( E'_k - E'_i (R')) / V(R')] dR'} , \label{sc5} \\
{\cal M}_{ik} = <\Phi_i e^{i (\mathbf{p}_i \mathbf{r})}
|\frac{\partial}{\partial R} -  \frac{\mathbf{v}_i +
\mathbf{v}_k}{2 V(R)} \mathbf{\nabla} |\Phi_k e^{i (\mathbf{p}_k
\mathbf{r})}> \;\;,  \label{sc6}
\end{eqnarray}
where the differential operators act only on the wavefunctions
$\Phi_i$ or $\Phi_k $, and $V(R)$ stands for the relative velocity
of the fragments. 

    The equations are to be solved with the initial condition
\beq C_i = \delta_{i1} \;\;\;\;\;\;\; for \;\;\  R = R_i\;\;,
\label{sc7} \eeq denoting that the muon is in the ground state at
the starting point $R_i$.

    Note, that for $R \to \infty$, the modified basis regains its
primordial orthogonality, as $p_i$ are the same for each state $i$
on one fragment.    Therefore, the amplitudes of decomposition
over the modified basis acquire the usual physical meaning. These
amplitudes must be used for the determination of physical
probabilities, rather than the amplitudes in the traditional basis
without the m.t.e.

Incorporation of the m.t.e. thus modifies the non-adiabatic matrix
element from \beq {\cal M'} = {{\partial}\over{\partial R}}
\;\;\;\; \label{sc8} \eeq to \beq {\cal M''} =
{{\partial}\over{\partial R}} + {{v_1 - v_2} \over {2V(R)}}
{{\partial}\over{\partial z}} \;\;\;,  \label{sc9} \eeq
 according to eq. (\ref{sc6}) with the momenta $p_n = const$.

      Equation (\ref{sc3}) poses a two-dimensional eigenvalue problem. It
has been solved numerically in ref.\cite{kar87} by making use of
the finite element method.   On the other hand, it should be noted
that using in eq. (\ref{sc1}) the asymptotic values of $p_n =
const$ is only justified in the limit $R \to \infty$. For small
$R$, where the relative velocity of the fragments is several times
smaller, the perturbation brought by the second terms in eqs.
(\ref{sc6}), (\ref{sc9}) to the non-adiabatic transition operator
is in principle excessively large. This difficulty was
successfully overcome \cite{jpg04}.

    The problem    (\ref{sc3})  has been solved in spheroidal coordinates
\beqa
\psi, \eta = (r_1 \pm r_2) R \label{sc4a} \\
1 \leq \psi < \infty,\;\;\;\;   -1 \leq \eta \leq 1 \;\;\;\; .
\label{sc4b} 
\eeqa 
It is noteworthy that recently, spheroidal coordinates were also
applied in refs. \cite{sphcr1,sphcr2}, where the dynamics was studied
of  light charged particle emission in spontaneous cold fission of
$^{252}$Cf. In these coordinates, the Laplacian becomes
\beq 
\Delta_{\psi ,\eta} = \frac{4}{R^2(\psi ^2 - \eta ^2)}
[\frac{\partial}{\partial \psi} (\psi ^2 - 1)
\frac{\partial}{\partial \psi} + \frac{\partial}{\partial \eta} (1
- \eta ^2) \frac{\partial}{\partial \eta} ] \;\;\;\; .
\label{sc5ab} 
\eeq

    A finite-differences method on a rectangular grid  (\ref{sc4b}) is used
to obtain a first approximation to the solution of the problem
(\ref{sc3}), generally a rather poor one. The final solution of the
problem is found by minimization of the Rayleigh-Ritz functional
of the problem (\ref{sc3}): \beq {\cal R}
 (\Phi) = \frac{    \int \int _G
[{\cal T} (\Phi) + \rho (\psi, \eta)\, (V_1 + V_2) \Phi ^2] d\psi
d\eta    }
{   \int \int _G   \rho (\psi, \eta)\, \Phi^2 d\psi d\eta }  \;\;\;\; ,  \label{sc6a} \\
\eeq where \beq {\cal T} (\Phi) = \frac{R}{4}[ (\psi ^2 - 1)
(\frac{\partial \Phi}{\partial \psi} )^2
+ (1 - \eta ^2) (\frac{\partial \Phi}{\partial \eta} )^2 ]  \;\;\;\; , \\
\eeq and \beq \rho (\psi, \eta)  = \frac{R^3}{8} (\psi^2 -
\eta^2)\;\;\;\; . \eeq The minimization is achieved by the
finite-element method. For realization of the method, all the area
$G$ of variables (\ref{sc4b}) is devided into N rectangular pieces
$g^j$. Within each element, a biquadratic  function is defined as
follows: \beq \phi_j (\psi, \eta) =  a_0^j + a_1^j \psi + a_2^j
\eta + a_3^j \psi \eta + a_4^j \psi^2  + a_5^j \eta^2 +  a_6^j
\psi^2 \eta + a_7^j \psi \eta^2 \;\;\;\; .     \label{sc8a} \eeq
In the basis of functions $\phi_j$, j = 1,...N, the problem
(\ref{sc6a}) is reduced to a generalized algebraic eigenvalue
problem $K \Phi = E M \Phi$, with the ``stiffness'' matrix $K$,
and $M$ the matrix of ``mass''. The problem
 is then solved by the inverse iteration method \cite{SIIM}.

Many results concerning the influence of the fission dynamics on 
the muon attachment probabilities were obtained by direct numerical integration 
of the time-dependent Schr\"odinger and Dirac equations on a 3-dimensional grid. 
This method and the results obtained are presented in the next sections.


\section{Theoretical Developments}

Because the nuclear excitation energy in muon-induced fission exceeds
the fission barrier height it is justified to treat the
fission dynamics classically (no barrier tunneling). For simplicity,
we describe the fission path by one collective coordinate $R$;
the classical collective nuclear energy has the form 
\begin{equation}
E_{\rm nuc} = \frac{1}{2} B(R) \dot R^2 + V_{\rm fis}(R) + E_\mu (R). 
\label{ecoll}
\end{equation}
We utilize a coordinate dependent mass parameter $B(R)$ \cite{OU93} and an empirical
double-humped fission potential $V_{\rm fis}(R)$ \cite{Ba74} which
is smoothly joined with the Coulomb potential of the fission fragments at
large $R$. The last term in Eq. (\ref{ecoll}) denotes the instantaneous muonic
binding energy which depends on the fission coordinate; this term will
be defined later. 

\begin{figure}[h!]
\vspace*{0.8cm}
\centerline{\includegraphics[scale=0.4]{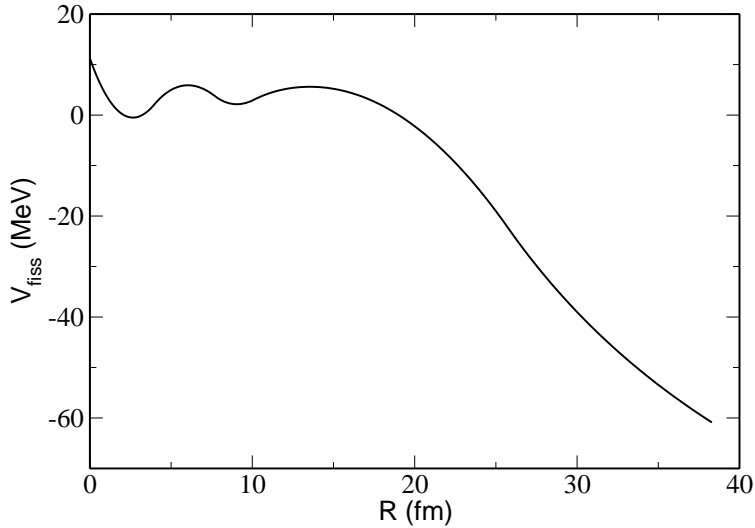}}
\vspace*{0.2cm}
\caption{\label{fispot} \protect\footnotesize Phenomenological fission potential for
$^{237}_{\ 93}Np$, for a mass asymmetry parameter $A_H/A_L=1.40$. The potential is
constructed from four parabolic sections and is smoothly joined with the Coulomb
potential of the fission fragments at large distance $R$.}
\end{figure}

To account for the nuclear energy dissipation between the outer fission barrier
and the scission point, we introduce a friction force which depends 
linearly on the velocity. In this case, the dissipation function $D$ is a simple
quadratic form in the velocity
\begin{equation}
\dot E_{\rm nuc}(t) = -2D = -f \dot R^2 (t) \label{frict}.
\end{equation}
The adjustable friction parameter $f$ determines the dissipated energy; it
is the only unknown quantity in the theory.

For the dynamical description of the muonic wavefunction during prompt
fission, the electromagnetic coupling between muon and nucleus $(-e \gamma
_\mu A^\mu)$ is dominant; the weak interaction is negligible. Because
of the nonrelativistic motion of the fission fragments the
electromagnetic interaction is dominated by the Coulomb interaction
\begin{equation}
A^0({\bf r},t) = \int d^3r' \frac{\rho_{\rm nuc}( {\bf r'},t)}
{| {\bf r} - {\bf r'} |}.    \label{vcoul}
\end{equation}
The muonic binding energy in the ground state of an actinide muonic atom
amounts to 12 percent of the muonic rest mass; hence non-relativistic
calculations, while qualitatively correct, are limited in accuracy. Several
theory groups have demonstrated the feasibility of such calculations
\cite{MO80,MW81,Ka97} which are based on the time-dependent Schr\"odinger equation
\begin{equation}
[ -\frac{\hbar^2}{2m} \nabla^2 -e A^0 ({\bf r},t) ] \ 
\psi ({\bf r},t) = i \hbar \frac{\partial}{\partial t} \ \psi ({\bf r},t) .
\end{equation}
In 1992, Oberacker et al.\cite{OU92,OU93} developed a numerical algorithm to
solve the relativistic Dirac problem on a three-dimensional Cartesian lattice.
The time-dependent Dirac equation for the muonic spinor wave function in the
Coulomb field of the fissioning nucleus has the form
\begin{equation}
H_{\rm D}(t) \ \psi ({\bf r},t) = i \hbar \frac \partial
{\partial t} \ \psi ({\bf r},t) \label{tdirac},
\end{equation}
where the Dirac Hamiltonian is given by
\begin{equation}
H_{\rm D}(t) =  -i \hbar c {\bf \alpha} \cdot \nabla + \beta mc^2
-e A^0 ({\bf r},t). \label{hdirac}
\end{equation}
Our main task is the solution of the Dirac
equation for the muon in the presence of a time-dependent external Coulomb
field $A^0({\bf r},t)$ which is generated by the fission
fragments in motion.
Note the coupling between the fission dynamics, Eq. (\ref{ecoll}), and
the muon dynamics, Eq. (\ref{tdirac}), via the
instantaneous muonic binding energy
\begin{equation}
E_\mu (R(t)) = \langle \psi ({\bf r},t) \mid H_{\rm D}(t) \mid
\psi ({\bf r},t) \rangle
\end{equation}
which depends on the fission coordinate; the presence of this
term increases the effective fission barrier height.


\section{Lattice Representation: Basis-Spline Expansion}

For the numerical solution of the time-dependent Dirac equation (\ref{tdirac})
it is convenient to introduce dimensionless space and time coordinates

$$
{\bf x} = {\bf r} / \lambdabar_c \ \ \ \ \lambdabar_c = \hbar /(m_\mu c)=1.87 fm
$$
\begin{equation}
\tau = t / \tau _c \ \ \ \ \tau _c= \lambdabar_c / c = 6.23 \times 10^{-24} s
\label{comptim}
\end{equation}
where $\lambdabar_c$ denotes the reduced Compton wavelength of the muon and
$\tau _c$ the reduced Compton time. For the lattice representation of the
Dirac Hamiltonian and spinor wave functions we introduce
a 3-dimensional rectangular box with a uniform lattice spacing $\Delta x$.
The lattice points are labeled $( x_\alpha, y_\beta, z_\gamma)$.

Our numerical algorithm is the Basis-Spline collocation method \cite{Uma91}.
Basis-Spline functions $B_i^M(x)$ are piecewise-continuous polynomials
of order $(M-1)$. These may be thought of as generalizations of the 
well-known ``finite elements'' which are B-Splines with $M=2$.

\begin{figure}[h!]
\vspace*{0.1cm}
\centerline{\includegraphics[scale=0.5,angle=-90]{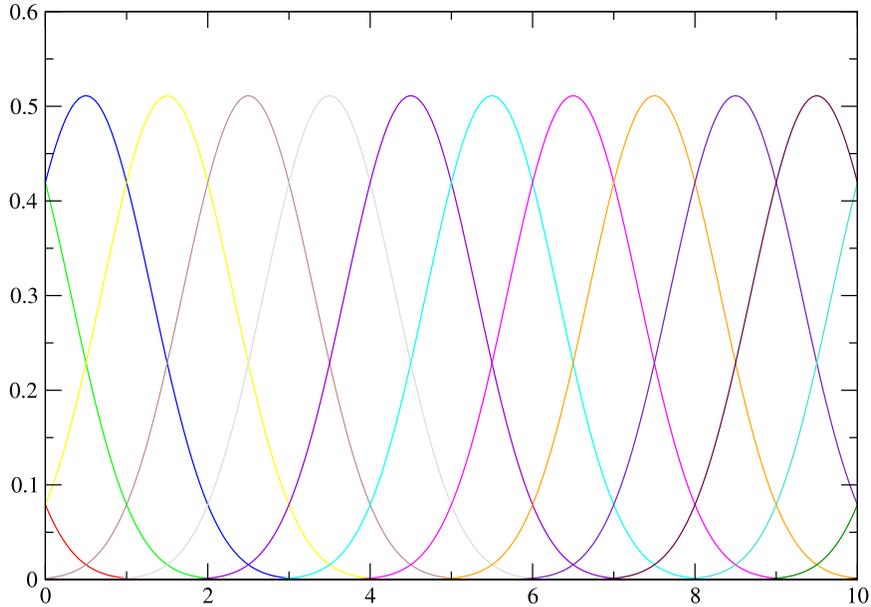}}
\vspace*{0.1cm}
\caption{\label{bspline_set} \protect\footnotesize Set of Basis Splines of order
$M=5$ using periodic boundary conditions.}
\end{figure}

There exists an extensive literature on Basis-Spline theory, developed by mathematicians
\cite{DeB78}. In 1991, Umar et al.\cite{Uma91} applied the method to the numerical
solution of problems in atomic and nuclear physics on a lattice; this paper
discusses the B-Spline collocation method, periodic and fixed boundary
conditions, the solution of various 1-D radial problems (Schr\"odinger, Poisson and
Helmholtz equations), and the solution of 3-D Cartesian problems (Poisson equation).
In a later paper in 1991, Umar et al. solved the nuclear static and time-dependent Hartree-Fock
equations on a 3-D lattice \cite{Uma91a}. During 1992-1995, Oberacker et al.\cite{OU92,OU93}
and Wells et al.\cite{WO95} applied the B-Spline collocation method to the
static and time-dependent Dirac equation which overcame the ``fermion doubling problem''
which one encounters with the finite-difference method. Very recently, we have used
both the B-Spline collocation and Galerkin methods for the solution of the
Hartree-Fock-Bogoliubov nuclear structure problem \cite{TOU03,OU03}.

To illustrate the B-Spline collocation method let us consider
a wave function which depends on one space coordinate $x$;
we represent the wave function on a finite spatial interval
as a linear superposition of B-Spline functions 
\begin{equation}
\psi(x_\alpha) = \sum_{i=1}^{N} B^M_i(x_\alpha) c^i .
\label{psialpha}
\end{equation}
In the Basis-Spline collocation method, local operators such
as the EM potential $A^0$ in Eq. (\ref{hdirac})
become diagonal matrices of their values at the grid points
(collocation points), i.e.\ $V(x) \rightarrow V_\alpha=V(x_\alpha)$.
The matrix representation of derivative operators is
more involved \cite{WO95}. For example, the first-derivative
operator of the Dirac equation has the following
matrix representation on the lattice
\begin{equation}
D_\alpha^\beta
\equiv  \sum_{i=1}^{N} B'_{\alpha i} B^{i \beta}\;,
\label{1der}
\end{equation}
where $B'_{\alpha i} = [dB_i^M(x) / dx] |_{x=x_\alpha}$.
Furthermore, we use the shorthand notation
$B_{\beta i}=B^M_i(x_\beta)$
for the B-spline function evaluated at the collocation point $x_\beta$,
and the inverse of this matrix is denoted by $B^{i \beta} =
[B^{-1}]_{\beta i}$.
Because of the presence of this inverse, the operator
$D_\alpha^\beta$ will have a nonsparse matrix representation.
In the present calculations we employ B-Splines of
order $M=5$. Eq. (\ref{psialpha}) can readily be generalized to three
space dimensions; in this case the four Dirac
spinor components $\psi ^{(p)}, p=( 1,\cdot \cdot \cdot,4)$
are expanded in terms of a product of Basis-Spline functions
\begin{equation}
\psi ^{(p)}( x_\alpha ,y_\beta ,z_\gamma ,t) = 
\sum\limits_{i,j,k}B^M_i(x_\alpha )B^M_j(y_\beta )B^M_k(z_\gamma )
c_{(p)}^{ijk}(t) ,
\end{equation}
i.e. the lattice representation of the spinor wave function
is a vector with $N = 4 * N_x * N_y * N_z$ complex components. Hence,
it is impossible to store $H_{\rm D}$ in memory because this would
require the storage of  $N^2$ complex double-precision numbers.
We must therefore resort to iterative methods for the solution of the matrix
equation which do not require the storage of $H_{\rm D}$.

We solve the time-dependent Dirac equation in two steps: first, we consider
the static Coulomb problem at time $t=0$, i.e. the muon bound to an
actinide nucleus
\begin{equation}
H_{\rm D}(t=0) \psi_{\rm gs} = E_{\rm gs} \psi_{\rm gs}.
\end{equation}
This static problem is solved by an iterative procedure (damped relaxation
method \cite{OU93,WO95}). The second part of our numerical procedure is
the solution of the time-dependent Dirac equation (\ref{tdirac})
by a Taylor-expansion of the propagator. For an infinitesimal time step
$\Delta t$ we find
\begin{equation}
\psi(t+\Delta t) = U(t+\Delta t,t) \psi(t) \approx
( 1 + \sum\limits_{n=1}^N \frac{(-i H \Delta t)^n}{n!} ) \psi(t) .
\end{equation}
We have thus reduced the original problem to a series of (matrix)$\times $%
(vector) operations which can be executed with high efficiency on vector or
parallel supercomputers without explicitly storing the matrix in memory.


\section{Discussion of Numerical Results}

In the following we present results for prompt fission of $^{237}_{\ 93}$Np
induced by the $E2:(3d \rightarrow 1s, 9.5 {\rm MeV})$ muonic transition.
All results reported here are for a 3-D Cartesian lattice of size
$L_x = L_y = 67$ fm and $L_z = 146$ fm with $N_x * N_y * N_z =
25 * 25 * 53$ lattice points with a uniform lattice spacing
$\Delta x = 1.5 \lambdabar_c = 2.8 fm$. Depending on the value of the
friction coefficient, we utilize between $1,200 - 1,900$ time steps with
a step size $\Delta t = 1.5 \tau_c = 9.3 \times 10^{-24}$ s.
The computer source code is written in Fortran 95. Typical production runs use
25 MB of memory and require 5.3 hours of CPU time on an Intel Pentium-4 (1.7GHz)
GNU/LINUX workstation. The same job takes 3.2 CPU hours on an IBM-SP2 supercomputer
at NERSC (serial run on one processor).

\begin{figure}[htb]
\vspace*{0.0cm}
\centerline{\includegraphics[scale=1.50]{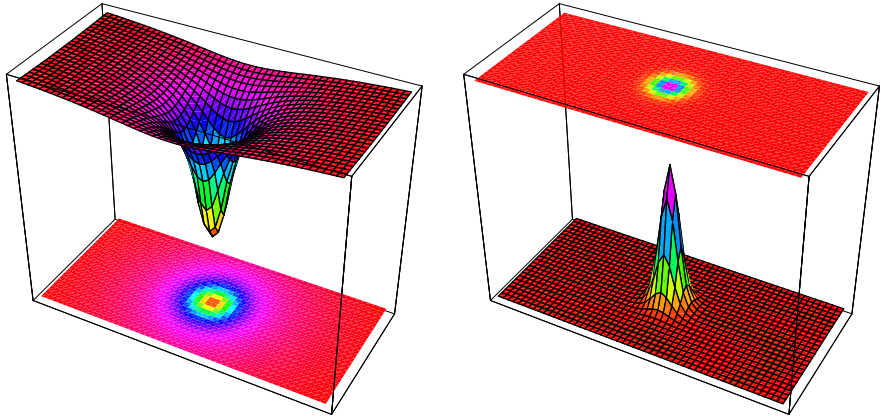}}
\vspace*{0.2cm}
\caption{\label{pl_t0} \protect\footnotesize Prompt muon-induced fission
 of $^{237}_{\ 93}$Np for a fission fragment mass
asymmetry $\xi = A_H / A_L = 1.10$ at $E^* = 9.5$ MeV. Shown is the Coulomb
interaction energy between the muon and the fissioning nucleus (left) and the corresponding muon
position probability density (right) at time $t = 0$ during fission.
Zero friction ($f=0$) has been assumed.}
\end{figure}

\begin{figure}[h!]
\vspace*{0.0cm}
\centerline{\includegraphics[scale=1.50]{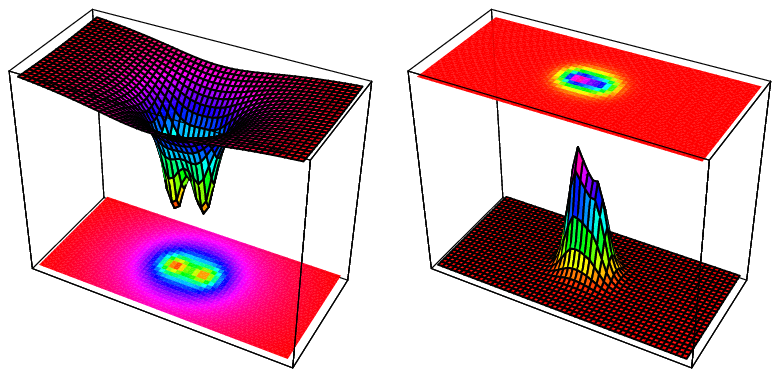}}
\vspace*{0.2cm}
\caption{\label{pl_t1} \protect\footnotesize Same as Fig.4, except at time step
$t = 6.5 \times 10^{-21}$ s.}
\end{figure}

\begin{figure}[h!]
\vspace*{0.0cm}
\centerline{\includegraphics[scale=1.50]{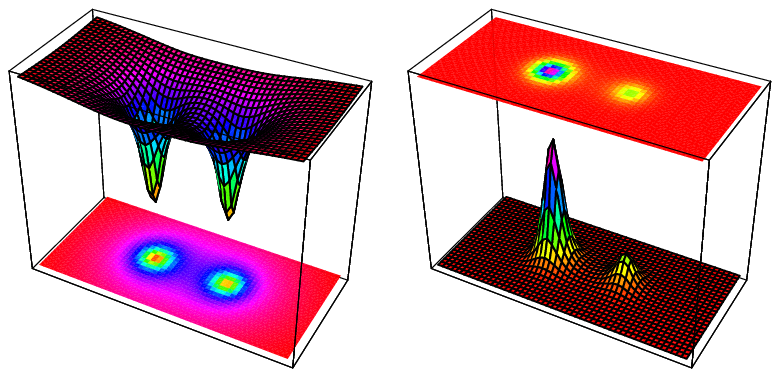}}
\vspace*{0.2cm}
\caption{\label{pl_t2} \protect\footnotesize Same as Fig.4, except at time step
$t = 8.4 \times 10^{-21}$ s.}
\end{figure}

\begin{figure}[h!]
\vspace*{0.0cm}
\centerline{\includegraphics[scale=1.50]{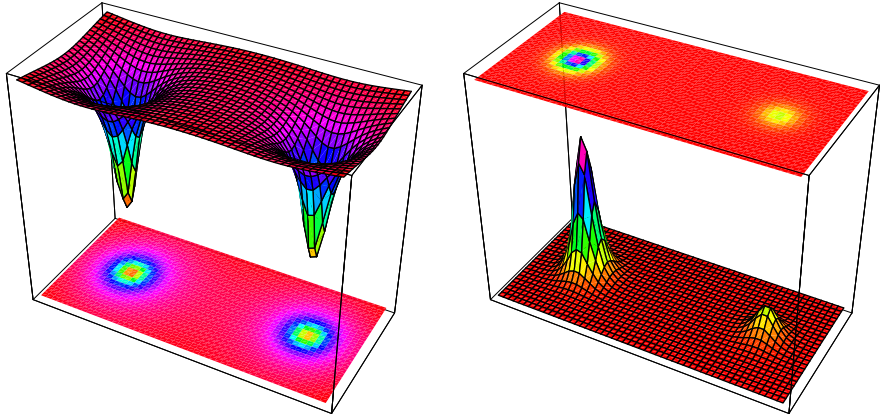}}
\vspace*{0.2cm}
\caption{\label{pl_t3} \protect\footnotesize Same as Fig.4, except at time step
$t = 1.1 \times 10^{-20}$ s.}
\end{figure}

Fig.\ \ref{pl_t0} shows the Coulomb interaction energy between the muon and
the neptunium nucleus (left) and the corresponding muon
position probability density (right) at time $t = 0$ during fission.
At this moment, the actinide nucleus is situated in its deformed ground state
minimum with quadrupole deformation $\beta_2=0.27$. 
Figures \ref{pl_t1},\ref{pl_t2},\ref{pl_t3} show the time-development
of the Coulomb interaction and the corresponding muon position probability
density during fission. The calculation has been done for an asymptotic
fragment mass asymmetry $\xi = A_H / A_L = 1.10$. In the last two time steps
the Coulomb potential wells of the two separated fission fragments are clearly
visible; the deeper well on the left is generated by the heavier fission
fragment. We observe a clear spatial correlation between the positions of the
potential minima and the muon probability density maxima.
As expected, the muon sticks predominantly to the heavy fragment (large bump
on the left in Fig.\ref{pl_t3}), but since the mass asymmetry in this case
is very small ($\xi = A_H / A_L = 1.10$) the muon attachment to the light fission
fragment (bump on the right) is relatively large. By integrating the
muon probability densities associated with the heavy and light fission
fragments at large internuclear distances, we can infer the muon attachment
probabilities to the heavy and light fragments, $P_H$ and $P_L$, respectively.
From Fig.\ref{pl_t3} we find $P_L=0.20$.

One might ask whether the muon will always remain bound during fission;
what is the probability for ionization? To investigate this question we
have plotted the muon position probability density on a logarithmic scale.

\begin{figure}[h!]
\vspace*{0.2cm}
\centerline{\includegraphics[scale=1.00]{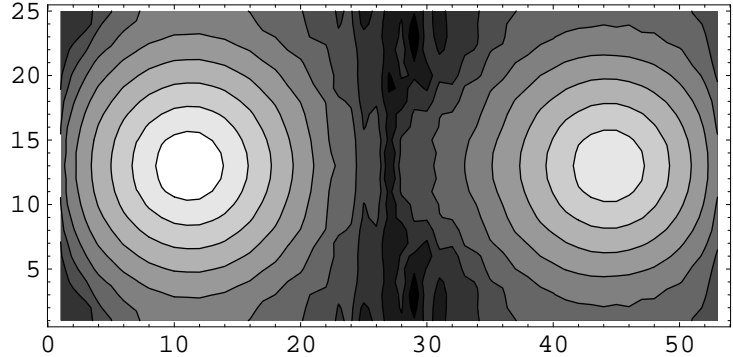}}
\vspace*{0.2cm}
\caption{\label{pl_log_ioniz} \protect\footnotesize Contour plot of the logarithm
 of the muon probability density at $ t = 1.1 \times 10^{-20}$ s shows
 no evidence of muon ionization.}
\end{figure}

In coordinate space, any appreciable muon ionization would show up as a
``probability cloud'' that is separating from the fission fragments and
moving towards the boundaries of the lattice. Fig.\ \ref{pl_log_ioniz} shows no
evidence for such an event in our numerical calculations. Hence, we conclude
that the probability for muon ionization $P_{\rm ion}$ is
substantially smaller than the muon attachment probability to the light
fission fragment which is always clearly visible in our logarithmic plots,
even at large mass asymmetry. From this we estimate that $P_{\rm ion}
< 10^{-4}$.

\begin{figure}[h!]
\vspace*{0.8cm}
\centerline{\includegraphics[scale=0.4]{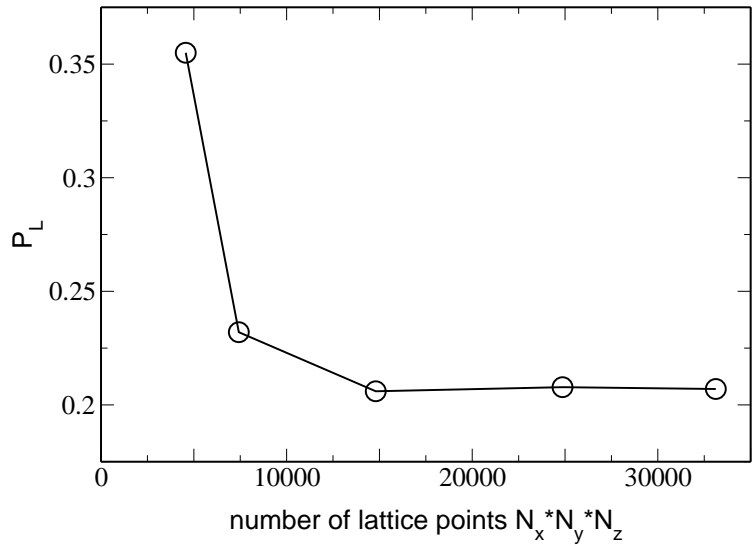}}
\vspace*{0.2cm}
\caption{\label{plconv} \protect\footnotesize Demonstration of numerical
 convergence of muon attachment probability for $^{237}_{\ 93}$Np.
 Calculations use a mass asymmetry of $1.10$ and zero friction.}
\end{figure}

As in all lattice calculations, we need to demonstrate convergence of our
results in terms of the lattice size and lattice spacing. Fig.\ \ref{plconv}
shows the asymptotic muon attachment probability $P_L$ as a function of
the total number of lattice points. In all of our production runs we use
a lattice with a total of $N_x * N_y * N_z = 25 * 25 * 53 = 33,125$
lattice points. It is apparent that the muon attachment probabilities
(and other related observables) have indeed converged.

\begin{figure}[h!]
\vspace*{1.0cm}
\centerline{\includegraphics[scale=0.40]{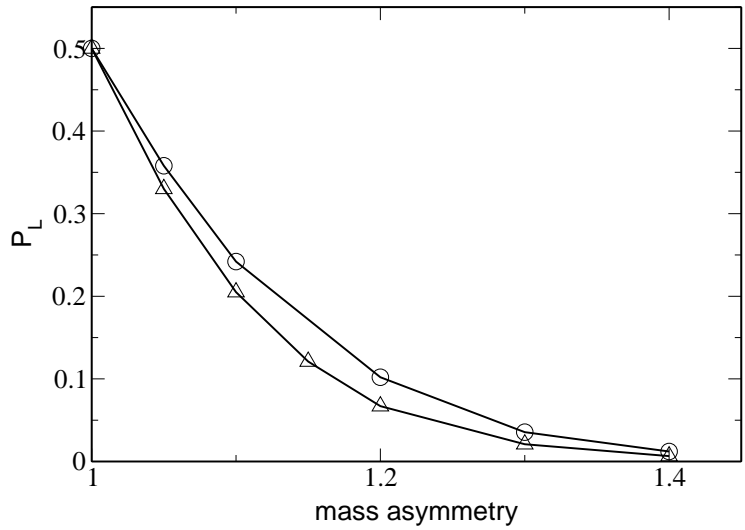}}
\vspace*{0.2cm}
\caption{\label{plmas} \protect\footnotesize Muon attachment to light fission
 fragment vs. fission fragment mass asymmetry.
upper curve: $^{238}_{\ 92}$U, lower curve: $^{237}_{\ 93}$Np. The calculations
assume zero friction.}
\end{figure}

Fig.\ \ref{plmas} shows that $P_L$ depends strongly on the fission fragment
mass asymmetry. This is easily understood: for equal fragments we must have
$P_L=0.5$, and for large mass asymmetry it is energetically favorable for
the muon to be bound to the heavy fragment, hence $P_L$ will be small.

\begin{figure}[h!]
\vspace*{1.0cm}
\centerline{\includegraphics[scale=0.4]{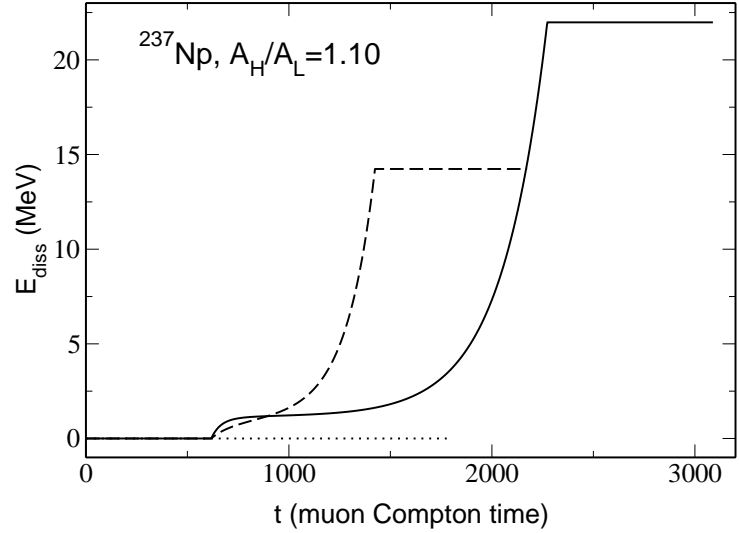}}
\vspace*{0.2cm}
\caption{\label{ediss} \protect\footnotesize Energy dissipated during fission, for
friction parameters $f=0$, $f=200$ and $f=500$.}
\end{figure}

Fig.\ \ref{ediss} shows the nuclear energy dissipation (in form of neutron-
and $\gamma $-emission) as a function of time; in our model, friction is
confined to the region between the outer fission barrier and the scission
point; for friction parameters $f=200$ and $f=500$, we obtain total
dissipated energies $E_{diss}=14.2$MeV and $22.0$MeV, respectively.

\begin{figure}[h!]
\vspace*{0.8cm}
\centerline{\includegraphics[scale=0.40]{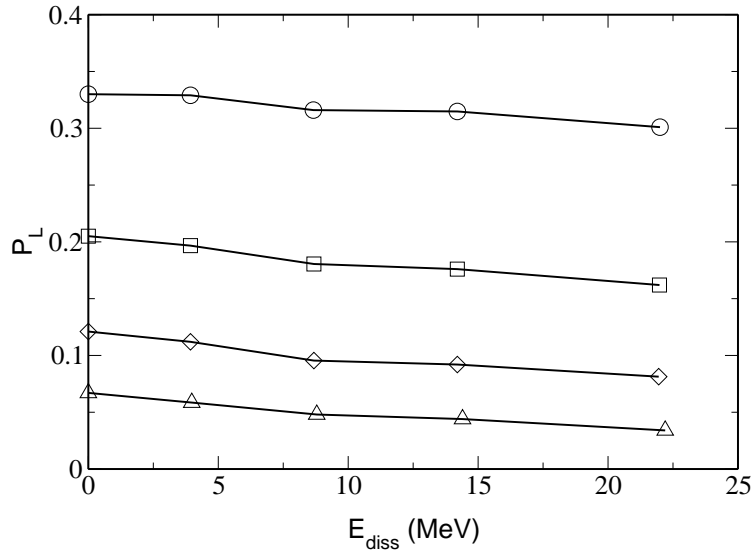}}
\vspace*{0.2cm}
\caption{\label{pl_ediss} \protect\footnotesize Muon attachment probability
 to the light fission fragment as
function of nuclear energy dissipation for $^{237}_{\ 93}$Np. Results are
shown for fragment mass asymmetries $\xi=1.05$ (upper curve), $1.10,\ 1.15$,
and $1.20$ (lower curve).}
\end{figure}

In Fig.\ \ref{pl_ediss} we examine the dependence of $P_L$ on the dissipated
nuclear energy, $E_{\rm diss}$, during fission. In our model, friction
takes place between the outer fission barrier and the scission point.
When the dissipated energy is computed from equation (\ref{frict})
we find an almost linear dependence of the muon attachment probability on
$E_{\rm diss}$; unfortunately, this dependence is rather weak.

We would like to point out that the theoretical values for $P_L$ obtained
in this work are smaller than those reported in our early calculations
\cite{OU92,OU93}. There are two reasons for this: (a) the size of
the lattice and (b) the lattice representation of the first derivative
operator in the Dirac equation. Because of constraints in the amount
of computer time available to us we utilized a smaller cubic lattice
in our prior calculations \cite{OU93} with $N_x * N_y * N_z = 29^3$
lattice points. In recent years, we were able to increase the size of the lattice
substantially, in particular in fission ($z$-) direction (see Fig.\ \ref{plconv}).
Regarding the lattice representation of the first
derivative operator, Eq. (\ref{1der}), in the Dirac equation: in 
ref. \cite{OU92,OU93} we utilized a combination of forward and backward
derivatives for the upper and lower spinor wave function components; after
extensive testing of Coulomb potential model problems with known analytical
solutions we have found that the symmetric derivative operator
provides a more faithful lattice representation. The results reported
here and in ref. \cite{OU98} have been obtained utilizing the symmetric
derivative prescription.


\section{Comparison of Theory with Experiment}

Prompt muon-induced fission was first observed experimentally
by Diaz et al. \cite{Di63}. More recent experiments by Ahmad et al. \cite
{AB80} yield a total fission probability per muon stop $P_f=0.068$ for
$^{238}U$ and a ratio $P_f$(prompt)/$P_f$(delayed)$=0.089$.

There are only few experimental data on muon attachment available for
comparison with our theory.
Schr\"oder {\it et al.} \cite{SW79} measured for the first time
mean lifetimes of muons bound to fission fragments of several
actinide nuclei. The muon decays from the K-shell of the muonic atom
through various weak interaction processes at a characteristic rate
$\lambda = \lambda_0 + \lambda_c$, where $\lambda_0 =
(2.2\times 10^{-6}s)^{-1}$ is the free leptonic decay rate for
the decay process $\mu^- \rightarrow e^- + \bar{\nu_e} + \nu_{\mu}$ and
$\lambda_c$ denotes the nuclear capture rate; $\lambda_c$ depends upon the
charge and mass of the fission fragment. From the observed lifetime $\tau _\mu
=1.30\times 10^{-7}s$ Schr\"oder {\it et al.} estimated an upper limit
for the muon attachment probability $P_L \le 0.1$. It must be
emphasized that this number represents an integral over the whole fission
mass distribution and, hence, cannot be directly compared to the
numbers given in Fig. \ref{pl_ediss}.

The most complete experiments have been carried out by Risse {\it et al.}
\cite{RB91} using the muon beam of the $\pi E3$ channel at the Paul
Scherrer Institute (PSI) in Switzerland. For this purpose, a fission
chamber has been inserted into the electron spectrometer SINDRUM I.
The incident muons are detected by a scintillation counter. An event
is defined by a $(\mu^-, f_1 f_2 e^-)$ coincidence where the fission
fragments are observed in prompt and the muon decay electrons in delayed
coincidence with respect to the incident muon. The magnetic field of the
electron spectrometer allows for a reconstruction of the
electron trajectories. Thus, it is possible to determine whether the muon
decay electrons originate from the heavy or the light fission fragment.

For several mass bins of the light fission fragment,
muon attachment probabilities $P_L$ have been measured; the experimental
data are given in Table \ref{exptheo}. It should be emphasized that the
mass bins are relatively broad. Because the theoretical values for $P_L$
depend strongly on the mass asymmetry it is not justified to assume that
$P_L$ remains constant within each experimental mass bin.
Instead, to allow for a comparison between theory and experiment,
we have to multiply the theoretical $P_L$ values in Fig.\ \ref{pl_ediss}
with a weighting factor that accounts for the measured relative mass
distribution \cite{RB91} of the prompt fission events within this mass bin.
We subsequently integrate the results over the sizes of the experimental
mass bins.
Due to the relatively low excitation energy in muon-induced fission,
the fission mass distribution exhibits a maximum at $\xi = A_H / A_L = 1.4$
and falls off rather steeply for values larger or smaller than the maximum.
This means that the large values of $P_L \approx 0.5$ at or near
fission fragment symmetry $\xi=1.0$ will be strongly suppressed.
The resulting theoretical values for $P_L$ are given in the last column of
Table \ref{exptheo}. It is apparent that our theory agrees rather well
with experiment. Because of the size of the error bars in the experiment
and because of the weak dependence of the theoretical values of $P_L$ on
the dissipated energy, it is not possible to extract very precise information
about the amount of energy dissipation.

\begin{table}[!t]
\caption{\protect\footnotesize Muon-attachment probabilities to the light
fission fragment, $P_L$, for $^{237}{\rm Np}(\mu^-,f)$. Exp. data are
taken from ref.\cite{RB91}.
\label{exptheo}}
\vspace{0.2cm}
\begin{center}
\footnotesize
\begin{tabular}{|c|c|c|c|}
\hline
{mass bin $A_L$} &\raisebox{0pt}[13pt][7pt]{mass asymmetry} &
\raisebox{0pt}[13pt][7pt]{$P_L$(exp)} &{$P_L$(theo)}\\
\hline
 & & & \\
$118.5 \rightarrow 111.5$   & $1.000 \rightarrow 1.126$
& $(25.5 \pm 8.5) \times 10^{-2}$
& $26.0 \times 10^{-2}, E_{\rm diss}=0 {\rm MeV}$ \\
    &    &    & $22.3 \times 10^{-2}, E_{\rm diss}=22 {\rm MeV}$ \\
 & & & \\
\hline
 & & & \\
$111.5 \rightarrow 104.5$   & $1.126 \rightarrow 1.268$
& $(9.7 \pm 2.6) \times 10^{-2}$
& $6.62 \times 10^{-2}, E_{\rm diss}=0 {\rm MeV}$ \\
    &    &    & $3.51 \times 10^{-2}, E_{\rm diss}=22 {\rm MeV}$ \\
 & & & \\
\hline
\end{tabular}
\end{center}
\end{table}

From a comparison of our theoretical result for the mass bin
$A_L = 118.5 \rightarrow 111.5$ with the measured data we extract
a dissipated energy of order $10$ MeV for $^{237}$Np while the second
mass bin $A_L = 111.5 \rightarrow 104.5$ is more compatible with zero
dissipation energy. We like to point out that the value $E_{\rm diss}=10$ MeV
agrees with results from other low-energy fission
measurements that are based on the odd-even effect in the charge yields
of fission fragments \cite{Wa91}. In addition to $^{237}$Np
we have also studied muon-induced fission of $^{238}$U; the results
for muon attachment are very similar (see Fig.\ \ref{plmas}).


\section{Outlook: calculation of dissipation in fission}

By making use of the commutators
$[\frac{\partial}{\partial R}, H]$  , $[\frac{\partial}{\partial
R}, z]$
one can express the non-adiabatic matrix elements (\ref{sc9}) in
terms of the mean-field derivatives and the energy differences of
the initial and final energies:

\beq {\cal M''}_{ik} = \frac{1}{\Delta E_{ik}} \left ( {{d}\over{d
R}} + {{v_1 - v_2} \over {2V(R)}} {{\partial}\over{\partial z}}
\right ) V(\vec{r}) \;\;\;. \label{sc9d} \eeq
 $V(\vec{r})$ in eq. (\ref{sc9d}) is the mean field potential. 
In the two-center model it can be written as follows:

 \beqa
 V(\vec{r}) \, = \,   \left\{
            \begin{array}{l@{\hspace{2cm}}l}
                     V_1(|\vec{r}-\vec{R_1}|)
                &  \mbox{for } z<0  \\
                     & \label{sc19} \\
                  V_2(|\vec{r}-\vec{R_2}|)
                &  \mbox{for } z\geq 0  \\
            \end{array}
   \right.
\eeqa

As the derivative is essentially non-zero at
the nuclear surface, this is in the spirit of the well-known
semi-classical wall-and-window model, where dissipation arises
from the interaction with the walls of a fissile nucleus.

We note that the change of the mean potential 
(\ref{sc19}) during separation of the fragments is due to two reasons: 
One is the change of the potential due to its radial dependence 
through $ V_i(|\vec{r}-\vec{R_i}|)$. The corresponding contribution 
to the matrix element (\ref{sc9d}) cancels \cite{BO2000} due to the equality
\beq \frac {\partial} {\partial R} + \frac {\partial}{\partial z}
\equiv 0 \;\;, \eeq which can be obtained by direct calculation.

The other contribution is due to ``breathing'' of each of the
nascent fragments, which arises from the change of its nuclear
radius $R_0$: \beq \frac {\partial} {\partial R_i}
V_i(|\vec{r}-\vec{R_i}|) =
 \frac {\partial V_i(|\vec{r}-\vec{R_i})} {\partial R_0}
\frac {\partial R_0} {\partial R}  \;\;\;. \label{sc20} \eeq

In the form (\ref{sc20}), the perturbation Hamiltonian was applied
for calculating
the energy dissipated during saddle~to~scission descent
\cite{BO2000} within the two-center harmonic oscillator model. The
calculated amount turns out to be rather small (about 1 MeV)
which gives evidence for a rather undamped motion for this kind of
friction. The result is in agreement with what was said in the
Introduction about this mechanism of nuclear friction. It also
agrees with the conclusion drawn from comparison of the calculated
muon attachment probabilities with experiment.


\section{Conclusions}

In this paper, we study the process of prompt muon-induced fission
of actinide nuclei. The nuclei are excited by radiationless transitions (inverse internal
conversion). For example, in $^{238}_{\ 92}$U the $E1: (2p \rightarrow 1s)$ and the
$E2: (3d \rightarrow 1s)$  muonic transitions result in excitation of the
nuclear giant dipole and giant quadrupole resonances, respectively, which act as 
doorway states for fission. It is very probable that the ($3d\rightarrow 1s, 9.5MeV$)
transition in the muonic atom will be dominant for muon-induced fission, because the transition
energy is very close to the peak of the $T=0$ giant quadrupole resonance
of the nucleus ($E_{GQR}=9.9MeV$). By contrast, there is a mismatch between the
($2p\rightarrow 1s, 6.6MeV$) muonic transition energy and the center of the
giant dipole resonance which is located at $E_{GDR}=12.8MeV$.
Because the muon lifetime is long compared to the timescale of prompt nuclear
fission, the motion of the muon in the Coulomb field of the fissioning nucleus
may be utilized to learn about the dynamics of fission.

We have studied the dynamics of a muon bound to a fissioning
actinide nucleus in two different ways: first, using the non-relativistic
time-dependent Schr\"odinger equation, the Born-Oppenheimer expansion method
was utilized. This method relies on a set of quasimolecular wavefunctions,
and the fission dynamics is described classically.

In the second method, we have solved the relativistic time-dependent Dirac
equation on a 3-D Cartesian lattice using Basis-Spline expansion
techniques. (The 3-D lattice makes it possible to investigate nuclear fission
shapes that are not axially symmetric. However, in the present study we
always assume axial nuclear symmetry, for simplicity). The principal advantage of the
coordinate lattice method is that it does not depend
on the validity of an ``atomic'' or ``molecular'' wavefunction basis set.
Furthermore, the method accounts not only for transitions between bound muonic states,
but is also describes the possible ionization of the muon during fission:
in coordinate space, any appreciable muon ionization would show up as a
``probability cloud'' that is separating from the fission fragments and
moving towards the boundaries of the lattice. Fig.\ \ref{pl_log_ioniz} shows no
evidence for such an event in our numerical calculations. 

Our time-dependent Dirac equation calculations predict a strong mass asymmetry
dependence of the muon attachment probability $P_L$ to the light fission
fragment; this feature is in agreement with experimental data. The theory also predicts
a (relatively weak) dependence of $P_L$ on the dissipated energy. By
comparing our theoretical results to the experimental data of
ref. \cite{RB91} we extract a dissipated energy of order $0-10$ MeV for
$^{237}$Np (see Table 1). Using the dissipation function defined in
Eq. (\ref{frict}), the $10$ MeV value corresponds to a fission time delay from
saddle to scission of order $2 \times 10^{-21}$ s.


\section*{Acknowledgements}

This work has been partially supported by the U.S. Department of Energy under grant No.
DE-FG02-96ER40963 with Vanderbilt University. Some of the numerical calculations
were carried out with the IBM-RS/6000 SP supercomputer (``Seaborg'') at the National Energy Research
Scientific Computing Center which is supported by the Office of Science of the
U.S. Department of Energy.


\end{document}